\documentclass[12pt,a4paper,notitlepage]{article}

\usepackage{amssymb}
\usepackage{amsmath}
\usepackage{amsfonts}

\begin{document}

\title{Quantum Computational Representation of the Bosonic Oscillator}
\author{George Jaroszkiewicz and Jason Ridgway-Taylor \\
School of Mathematical Sciences, University of Nottingham, \\
University Park, Nottingham NG7 2RD, UK}
\maketitle

\begin{abstract}
By the early eighties, Fredkin, Feynman, Minsky and others were exploring
the notion that the laws of physics could be simulated with computers.
Feynman's particular contribution was to bring quantum mechanics into the
discussion and his ideas played a key role in the development of quantum
computation. It was shown in 1995 by Barenco et al that all quantum
computation processes could be written in terms of local operations and CNOT
gates. We show how one of the most important of all physical systems, the
quantized bosonic oscillator, can be rewritten in precisely those terms and
therefore described as a quantum computational process, exactly in line with
Feynman's ideas. We discuss single particle excitations and coherent states.
\end{abstract}

\section{Introduction}

The harmonic oscillator is one of the most important dynamical systems in
mathematical physics, providing a basis for the description of many
fundamental physical concepts. These range from normal mode analysis in
classical many-body theory, the coulombic interaction in quantum mechanics
\cite{CORNISH-84}, the algebra of angular momentum operators \cite%
{SCHWINGER-65}, supersymmetric quantum mechanics \cite{MEJIA+PLEITEZ-02},
coherent states in quantum optics, free particle states and perturbation
theory expansions in relativistic quantum field theory, excitations of
relativistic strings and many other examples. The oscillator also provides a
framework for the description of bosons, which is part of the focus of this
work. It is a completely solvable system and has been described from
numerous points of view in both the classical and quantum formalisms.

In this article our aim is to show how the quantized one-dimensional
harmonic oscillator can be described in quantum computational terms. This
requires the introduction of a suitable Hilbert space, a \emph{quantum
register}, over which the computational processes are carried out. It turns
out that, because bosons have no upper limit to their excitation levels,
this involves a tremendous amount of mathematical redundancy. The Hilbert
space dimension of the quantum register introduced is far bigger than that
actually needed to describe a quantized boson, most of the states in the
register being what we call \emph{transbosonic}. It is in fact
non-separable, i.e., has no denumerable basis, whereas the conventional
Hilbert space used to describe the quantum harmonic oscillator is
denumerable. We will show exactly how bosonic states are restricted to a
tiny subspace of the register and remain there during the quantum
computational processes involved in the quantum register approach.

The motivation for taking this approach arises from a combination of
circumstances. In the nineteen fifties, Ulam and von Neumann began to
discuss computational models known as cellular automata, in which simple
rules of computation applied to systems with many degrees of freedom could
produce complex patterns of behaviour. By the nineteen eighties, Fredkin,
Feynman \cite{FEYNMAN-82}, Minsky \cite{MINSKY-82} and others were
speculating on the possibility of describing the laws of physics and the
universe in terms of cellular automata and computation. Underlying their
ideas was a dissatisfaction with the conventional description of physics
based on continuous space and time. In their work, the number of
computational degrees of freedom was generally considered very large but
countable. This has found resonance recently in a number of directions, such
as the spin network approach to quantum gravity and the holographic
principle, which asserts that the information content within a region of
spacetime is limited by how much information can be packed over its defining
surface, which can hold no more than about $10^{69}$ bits per square metre.

Feynman's approach is notable because of his interest in applying quantum
principles to computation. In due course, the field of quantum computation
emerged as a major theoretical and experimental field of research, and
Feynman's influence on this has been well-documented \cite{HEY-99}. At
present, the practical difficulties involved in the construction of an
actual quantum computer have focused most attention on systems consisting of
a relatively small number of quantum bits. This is in contrast to the
thinking behind the work of Fredkin, Minsky, Wolfram and others, who
envisaged the universe as consisting of a vast number of degrees of freedom.

Our long-term interest is in seeing how far the principles of quantum
computation can be extended to cover much larger systems, such as the
universe, in line with the ideas of the pioneers. However, it is important
to show that these ideas can also apply to real physical systems, such as
those encountered in the laboratory. A particularly good area for
discussion in this direction is quantum optics, for two reasons. First,
experiments in quantum optics are particularly clean both theoretically
and experimentally. Secondly, devices such as beam-splitters, mirrors,
Wollaston prisms, lenses and phase-shifters can be linked together in a
modular way to build up more complex apparatus, such as Mach-Zender
interferometers. It should also be possible to couple many beam-splitters
\cite{ELITZUR-01} and Mach-Zender interferometers together to form much
larger systems. This is very similar to the idea of linking many single
serial processors\ to form a parallel processing computer. We imagine the
process of linking more and more pieces of apparatus being continued to
such an extent and over such distance scales that the result begins to
look as if the universe itself is one gigantic quantum experiment. Quantum
optics provides one scenario in which Feynman's vision of a universe, or at
least a substantial part of it, and its associated laws, as being described
in quantum computational terms looks reasonable and useful. The link with
these ideas and the harmonic oscillator arises quite naturally because
photons are bosons and have many of the characteristics of oscillators.

The plan of this paper is as follows. First we discuss some basic concepts
involved in our formalism. We start with the notion of a \emph{logic function%
}, which is a function from any given domain into the discrete set $\{0,1\}$%
, relating it to classical bits and classical register states. This puts us
in a position where we can discuss classical computation and the CNOT gate.
This is followed by a discussion of quantization and quantum computation,
where we introduce a quantum variant of the CNOT discussed in the classical
case. This quantum version of the CNOT is required for our discussion of the
quantized bosonic oscillator which then follows. We show how the standard
oscillator can be seen as a form of quantum computation in an infinite rank
quantum register, i.e., an infinite collection of quantum bits. We conclude
with a review of the quantum register description of coherent states, which
are important in many realistic situations.

On the matter of notation, we distinguish states and operators involved in
the standard quantum theory and their analogues in the quantum register
description using the following conventions. States described by standard
Hilbert space vectors will always be represented using angular Dirac
bra-ket notation, such as $|\Psi \rangle $, $\langle \phi |$, etc.,
whereas states represented by elements of a quantum register will be
described by a modified form of this notation, replacing the angular
bracket with a round bracket, i.e., $|\psi )$, $(f|$, etc. The quantum
register analogue of an operator $\hat{O}$ in the standard theory will be
denoted by $\mathbb{O}$, with one or two exceptions.

\section{Logic functions, bits and states}

To explain our approach fully, it will be helpful to review some basic
concepts associated with classical logic and computation.

Let $X$ be a discrete indexing set for another set $Y$, the latter set
consisting of objects having certain properties in which we are interested.
Each element $x$ of $X$ will be called a \emph{site} and serves as a label
for a corresponding unique element $y(x)$ of $Y$. The number $N$ of elements
in $X$ will be assumed finite for now, but taken to infinity when we apply
our ideas to the bosonic oscillator. For convenience, we shall label the
elements of $X$ by an integer running from zero to $N-1$, i.e., $X=\left\{
x_{0},x_{1}.\ldots ,x_{N-1}\right\} $.

Suppose we have asked a particular question $f$ of the object $y(x)$ at site
$x$, this question having only one of two possible answers, either \emph{yes}
($\equiv $ true) or \emph{no} ($\equiv $ false). If the answer is \emph{yes}
we will associate the number $f(x)=1$ with $x$; otherwise we will associate
the number $f(x)=0$.

Assuming we can ask the same question of all elements in $Y$, we now define
the \emph{logic function} $f(X)$ to be a real valued function from $X$ into
the discrete set $\left\{ 0,1\right\} $, with its value $f(x)$ at each site $%
x$ in $X$ defined by the answer to the question $f$ asked of the
corresponding $y(x)$ in $Y$.

If we had asked some other question $g$ of $y(x)$, we would have obtained an
answer $g(x)$ at site $x$ which could in principle differ to the answer $%
f(x) $ to question $f$. Each question therefore defines a particular logic
function over $X$. We shall denote the set of all distinct logic functions
over $X$ by $\mathbb{L}(X)$. If two questions have the same logic function
then we shall regard these questions as synonymous, i.e. equivalent.
Otherwise, we shall regard these questions as distinct. For finite $N$, it
is easy to see that there are exactly $2^{N}$ distinct logic functions in $%
\mathbb{L}(X)$. Two important logic functions which occur in every set of
logic functions are the \textquotedblleft yes\textquotedblright\ function $%
Yes$, which satisfies the property $Yes\left( x\right) =1$ for every site $x$
in $X$, and the \textquotedblleft no\textquotedblright\ function $No$, which
satisfies the property $No\left( x\right) =0$ for every site $x$ in $X$.

In classical computation, we are generally interested in various sorts of
maps involving logic functions. We define a \emph{type} $\left( r,s\right) $
\emph{gate} to be a map from the Cartesian product $\mathbb{L}^{r}\left(
X\right) \equiv \mathbb{L}\left( X\right) \times \mathbb{L}\left( X\right)
\times \ldots \times \mathbb{L}\left( X\right) \;(r$ copies of $\mathbb{L}%
\left( X\right) $) into the Cartesian product $\mathbb{L}^{s}\left( X\right)
$. A \emph{binary gate} is a type $\left( 2,1\right) $ gate and a \emph{rank-%
}$n$\emph{\ gate} is a type $(n,n)$ gate.

Binary gates can be defined via truth tables, in which ``$yes"=1=T=``true"$
and $``no"=0=F=``false".$ Three important binary gates are:

\begin{enumerate}
\item[i)] the \textquotedblleft \emph{AND}\textquotedblright\ binary gate $%
\wedge $, defined by
\begin{equation}
\left( f\wedge g\right) \left( x\right) \equiv f\left( x\right) \wedge
g\left( x\right) ,\;\;\;\;\;\forall x\in X,
\end{equation}%
where $\wedge $ denotes the logical \textquotedblleft and\textquotedblright\
applied to the truth value associated with $f\left( x\right) $ and $g\left(
x\right) ;$

\item[ii)] the \emph{``OR'' }binary gate $\vee $, defined by
\begin{equation}
\left( f\vee g\right) \left( x\right) \equiv f\left( x\right) \vee g\left(
x\right) ,\;\;\;\;\;\forall x\in X,
\end{equation}
where $\vee $ is the logical ``inclusive or'';

\item[iii)] the \textquotedblleft $XOR$\textquotedblright\ binary gate,
defined by
\begin{equation}
\left( f\oplus g\right) \left( x\right) \equiv f\left( x\right) \oplus
g\left( x\right) ,\;\;\;\;\;\forall x\in X,  \label{XOR}
\end{equation}%
where $\oplus $ denotes addition modulo two. This is also known as
\textquotedblleft exclusive or\textquotedblright .\
\end{enumerate}

Separately, each of these three binary gates are commutative and, when
generalized to type $(3,1)$ gates, associative, but this is not true of
other binary gates in general.

Associated with every logic function $f$ in $\mathbb{L}\left( X\right) $ is
another logic function $\tilde{f}$, known as its \emph{negation}, or logical
dual, defined by the statement
\begin{equation}
f\oplus \tilde{f}=Yes,
\end{equation}
which can be solved to give
\begin{equation}
\tilde{f}=f\oplus Yes.
\end{equation}

\section{\textbf{Classical bits}}

We will find it very useful to encode logic functions in terms of
two-component objects called \emph{bits}, analogous to classical (Pauli)
spinors. It should be kept firmly in mind, however, that despite
appearances, these objects are neither spinors, fermions, nor Grassmannian
variables, and have nothing to do with spin-half angular momentum. They are
simply a way of encoding elementary pieces of information, i.e., truth
values, in computational terms.

Given a logic function $f\in \mathbb{L}\left( X\right) $, define the \emph{%
classical bit} \emph{state }(henceforth abbreviated to \emph{bit state}) $%
\psi _{f}\left( x\right) $ \ at $x\in X$ as the two component object
\begin{equation}
\psi _{f}\left( x\right) \equiv \left[
\begin{array}{c}
f\left( x\right) \\
\tilde{f}\left( x\right)%
\end{array}%
\right] _{x}.
\end{equation}%
Here the subscript on the column vector reminds us that this particular bit
state refers to element $x$ and not to any other element of $X$. The set of
all possible distinct bit states at $x$ defines a classical bit
\emph{subregister}, denoted by $\mathcal{R}_{x}$. The
\emph{\textquotedblleft basis vectors\textquotedblright }
\begin{equation}
|0)_{x}\equiv \psi _{No}\left( x\right) =\left[
\begin{array}{c}
0 \\
1%
\end{array}%
\right] _{x},\;\;\;|1)_{x}\equiv \psi _{Yes}\left( x\right) =\left[
\begin{array}{c}
1 \\
0%
\end{array}%
\right] _{x},
\end{equation}%
provide a basis for $\mathcal{R}_{x},$ in that any classical bit state $\psi
_{f}\left( x\right) $ at $x$ can be written as the linear combination
\begin{equation}
\psi _{f}\left( x\right) =\tilde{f}\left( x\right) |0)_{x}+f\left( x\right)
|1)_{x},
\end{equation}%
using the standard rules for matrix multiplication and addition. However,
because the components $f\left( x\right) $ and $\tilde{f}\left( x\right) $
can take values one or zero only, $\mathcal{R}_{x}$ is not a vector space
over the reals.

The \emph{adjoint bit state} $\psi _{f}^{+}\left( x\right) $ at $x$ is
defined by
\begin{equation}
\psi _{f}^{+}\left( x\right) \equiv \left[
\begin{array}{cc}
f\left( x\right) & \tilde{f}\left( x\right)%
\end{array}%
\right] _{x}=f\left( x\right) {}_{x}(1|+\tilde{f}\left( x\right) {}_{x}(0|,
\end{equation}%
where
\begin{equation}
{}_{x}(0|\equiv \left[ 0\;1\right] {}_{x},\;\;\;{}_{x}(1|\equiv \left[ 1\;0%
\right] {}_{x}.
\end{equation}%
Using the rule%
\begin{equation*}
_{x}(i|j)_{x}=\delta _{ij},\ \ \ i,j=0,1,
\end{equation*}%
where $\delta _{ij}$ is the Kronecker delta, we find
\begin{equation}
\psi _{f}^{+}\left( x\right) \psi _{f}\left( x\right) =f^{2}\left( x\right) +%
\tilde{f}^{2}\left( x\right) =1,
\end{equation}%
i.e.,
\begin{equation}
\psi _{f}^{+}\psi _{f}=Yes,\;\;\;\;\;\forall f\in \mathbb{L}\left( X\right) .
\end{equation}

The \emph{anti-bit state} $\tilde{\psi}_{f}$ is defined by
\begin{equation}
\tilde{\psi}_{f}\left( x\right) \equiv {\sigma }_{x}^{1}\psi _{f}\left(
x\right) =\psi _{\tilde{f}}\left( x\right) ,
\end{equation}%
where
\begin{equation}
\sigma _{x}^{1}\equiv |0)_{x}(1|+|1)_{x}(0|=\left[
\begin{array}{cc}
0 & 1 \\
1 & 0%
\end{array}%
\right] _{x}
\end{equation}%
is one of the Pauli matrices. This matrix plays the role of the negation
operation and is called a \emph{flip}. We have the relations
\begin{equation}
\tilde{\psi}_{f}^{+}\psi _{f}=\psi _{f}^{+}\tilde{\psi}_{f}=No,\;\;\;\;\;%
\forall f\in \mathbb{L}\left( X\right) .
\end{equation}%
\qquad \qquad

The usefulness of this encoding is that we can readily express a number of
natural questions in bit terms. For a finite set, the cardinality $\#X$ of
the set $X$ is given by
\begin{equation}
\#X\equiv \sum_{x\in X}\psi _{f}^{+}\left( x\right) \sigma _{x}^{0}\psi
_{f}\left( x\right) =\sum_{x\in X}1,\;\;\;\forall f\in \mathbb{L}\left(
X\right) ,
\end{equation}%
where $\sigma _{x}^{0}$ is the $2\times 2$ identity operator at $x$. The
number $Y_{f}\left( X\right) $ of \textquotedblleft yes\textquotedblright\
values over $X$ relative to the logic function $f$ is given by
\begin{equation}
Y_{f}\left( X\right) \equiv \sum_{x\in X}\psi _{f}^{+}\left( x\right)
P_{x}^{1}\psi _{f}\left( x\right) =\sum_{x\in X}f\left( x\right) ,
\end{equation}%
where
\begin{equation}
P_{x}^{1}\equiv |1)_{x}(1|=\left[
\begin{array}{cc}
1 & 0 \\
0 & 0%
\end{array}%
\right] _{x}.
\end{equation}%
Likewise, the number $N_{f}\left( X\right) $ of \textquotedblleft
no\textquotedblright\ values over $X$ relative to the logic function $f$ \
is given by
\begin{equation}
N_{f}\left( X\right) \equiv \sum_{x\in X}\psi _{f}^{+}\left( x\right)
P_{x}^{0}\psi _{f}\left( x\right) =\sum_{x\in X}\tilde{f}\left( x\right) ,
\end{equation}%
where
\begin{equation}
P_{x}^{0}\equiv |0)_{x}(0|=\left[
\begin{array}{cc}
0 & 0 \\
0 & 1%
\end{array}%
\right] _{x}.
\end{equation}%
The difference $Y_{f}\left( X\right) -N_{f}\left( X\right) $ is given by
\begin{equation}
Y_{f}\left( X\right) -N_{f}\left( X\right) =\sum_{x\in X}\psi _{f}^{+}\left(
x\right) \sigma _{x}^{3}\psi _{f}\left( x\right) ,
\end{equation}%
where $\sigma _{x}^{3}$ is the third Pauli matrix
\begin{equation}
\sigma _{x}^{3}\equiv P_{x}^{1}-P_{x}^{0}=\left[
\begin{array}{cc}
1 & 0 \\
0 & -1%
\end{array}%
\right] _{x}.
\end{equation}

We will expand the algebraic structure of these operators in anticipation of
subsequent extension to quantum bits. At each site $x$, define the
\textquotedblleft transition operators\textquotedblright\
\begin{equation}
A_{x}\equiv |0)_{x}(1|,\;\;\;\;\;A_{x}^{+}\equiv |1)_{x}(0|.
\end{equation}%
Then at any given site, the operators $P^{0},P^{1}$, $A$, $A^{+}$ and the
zero operator $0$ can be multiplied together to give the closed algebra of
operators\textrm{\ }shown in Table $1$, where for example the product $%
P^{1}A=0$\ is read off from the intersection of the row labelled $P^{1}$ and
the column labelled $A$:

\begin{center}
\begin{tabular}[t]{|c|ccccccc|}
\hline
$^{\;}$ & ${P}^{0}$ & ${P}^{1}$ & ${A}$ & ${A}^{+}$ & ${\sigma }^{1}$ & ${%
\sigma }^{2}$ & ${\sigma }^{3}$ \\ \hline
${P}^{0}$ & $P^{0}$ & $0$ & $A$ & $0$ & $A$ & $iA$ & $-P^{0}$ \\
${P}^{1}$ & $0$ & $P^{1}$ & $0$ & $A^{+}$ & $A^{+}$ & $-iA^{+}$ & $P^{1}$ \\
${A}$ & $0$ & $A$ & $0$ & $P^{0}$ & $P^{0}$ & $-iP^{0}$ & $A$ \\
${A}^{+}$ & $A^{+}$ & $0$ & $P^{1}$ & $0$ & $P^{1}$ & $iP^{1}$ & $-A^{+}$ \\
${\sigma }^{1}$ & $A^{+}$ & $A$ & $P^{1}$ & $P^{0}$ & $\sigma ^{0}$ & $%
i\sigma ^{3}$ & $-i\sigma ^{2}$ \\
${\sigma }^{2}$ & $-iA^{+}$ & $iA$ & $-iP^{1}$ & $iP^{0}$ & $-i\sigma ^{3}$
& $\sigma ^{0}$ & $i\sigma ^{1}$ \\
${\sigma }^{3}$ & $-P^{0}$ & $P^{1}$ & $-A$ & $A^{+}$ & $i\sigma ^{2}$ & $%
-i\sigma ^{1}$ & $\sigma ^{0}$ \\ \cline{1-7}\cline{2-8}\cline{8-8}
\end{tabular}

\

Table 1. The product rule for projection, transition, and Pauli operators.
\end{center}

\section{\textbf{Classical logic registers}}

In the above, the set $X$ is treated as if it were a physical space over
which we have defined a local bit state $\psi _{f}\left( x\right) $ at each
site $x$. This is analogous to the description of a system of $N$ point
particles in Newtonian mechanics where at each instant of time each particle
is regarded as having a well-defined position in three-dimensional
(physical) space with some instantaneous velocity. A more sophisticated but
entirely analogous approach is to consider such an $N$-particle system as a
single point in a $6-N$ dimensional phase space. Likewise, an equivalent
description of our qubit system is to adopt a tensorial approach, where we
define the classical register state $|f)$ as the \textquotedblleft tensor
product\textquotedblright\
\begin{equation}
|f)\equiv \psi _{f}\left( x_{1}\right) \otimes \psi _{f}\left( x_{2}\right)
\otimes \ldots ,\;\;\;f\in \mathbb{L}\left( X\right) ,  \label{123}
\end{equation}%
the symbol $\otimes $ denoting the equivalent of the Kronecker product.
Here, the product contains one factor for every element of $X$. The ordering
of factors in this product is not significant at the formal level and what
really matters are the labels. Of course, once a matrix representation has
been chosen for such a state, the relative ordering of labels does become
important and must be kept fixed.

The number of distinct classical register states is the same as the
cardinality of $\mathbb{L}\left( X\right) $, i.e., $2^{N}$. The set of
distinct classical register states of the form $\left( \ref{123}\right) $
defines what we call the \emph{classical register} $\mathcal{R}_{X}$ over $X$
and this is the analogue of classical phase space. It is important to keep
in mind that, just as in the case of classical phase space, the
\textquotedblleft vector\textquotedblright\ addition of different elements
of the classical register has no classical interpretation, but there will be
a quantum interpretation of such addition. The cardinality $N$ of the
indexing set $X$ will be referred to as the \emph{rank} of the associated
register $\mathcal{R}_{X}$, in both classical and quantum versions. We shall
henceforth use the notation $\mathcal{R}^{N}$ to mean the rank-$N$ register $%
\mathcal{R}_{X}$. An infinite-rank register will be denoted by $\mathcal{R}%
^{\infty }$.

Each logic function $f$ can be mapped into a unique integer $n_{f}$ in the
interval $\left[ 0,2^{N}-1\right] $ by the rule
\begin{equation}
n_{f}=\sum_{i=0}^{N-1}f\left( x_{i}\right) 2^{i},  \label{smap}
\end{equation}%
and we may use the notation $|f)$ or $|n_{f})$ to denote the corresponding
register state when there is no ambiguity. We shall call this rule the \emph{%
computational map. }

Two of the states in the register $\mathcal{R}^{N}$ are particularly
important; the \emph{void state }$|0)$ is the state associated with the
\textquotedblleft No\textquotedblright\ logic function over $X,$ i.e.,%
\begin{equation}
|No)\equiv |0)=\left[
\begin{array}{c}
0 \\
1%
\end{array}%
\right] _{x_{1}}\otimes \left[
\begin{array}{c}
0 \\
1%
\end{array}%
\right] _{x_{2}}\otimes \ldots \otimes \left[
\begin{array}{c}
0 \\
1%
\end{array}%
\right] _{x_{N}},
\end{equation}%
and the \textquotedblleft fully excited\textquotedblright\ state, associated
with the \textquotedblleft Yes\textquotedblright\ logic function:
\begin{equation}
|Yes)\equiv |2^{N}-1)=\left[
\begin{array}{c}
1 \\
0%
\end{array}%
\right] _{x_{1}}\otimes \left[
\begin{array}{c}
1 \\
0%
\end{array}%
\right] _{x_{2}}\otimes \ldots \otimes \left[
\begin{array}{c}
1 \\
0%
\end{array}%
\right] _{x_{N}}.
\end{equation}%
The void state should not be thought of as either a \textquotedblleft
vacuum" state or as a \textquotedblleft ground state". The reason for this
will become evident when we show how to describe the harmonic oscillator in
quantum register terms. The ground state of the harmonic oscillator actually
corresponds to the first excited state of the void state. The void state is
an example of a \emph{transbosonic} state, defined below, and its
interpretation will be discussed in our next paper, when we describe actual
physics experiments via quantum registers.

Dual register states $(f|$ are defined by
\begin{equation}
(f|\equiv \psi _{f}^{+}\left( x_{N}\right) \otimes \ldots \otimes \psi
_{f}^{+}\left( x_{2}\right) \otimes \psi _{f}^{+}\left( x_{1}\right) ,
\end{equation}%
and then the \textquotedblleft inner product\textquotedblright\ $(f|g)$ is
defined in the obvious way:
\begin{equation}
(f|g)=\prod_{x\in X}\psi _{f}^{+}\left( x\right) \psi _{g}\left( x\right) .
\end{equation}%
This inner product has the property that
\begin{equation}
(f|g)=\delta _{f,g},
\end{equation}%
where $\delta _{f,g}$ is the logic function equivalent of the Kronecker
delta, i.e., takes value unity for $f=g$ and zero otherwise. The $2^{N}$
\textquotedblleft orthonormal\textquotedblright\ classical register states
in $\mathcal{R}^{N}$ form a convenient basis, known as the \emph{%
computational basis}, for the Hilbert space $\mathcal{H}^{N}$ associated
with the classical register $\mathcal{R}^{N}$ when we have quantized the
system.

\section{Classical computation and CNOT}

Before we can discuss quantum computation, we need to review some notions of
classical computation. Consider a classical system composed of two sites
denoted by $a$\ and $b,$ i.e., $X=\left\{ a,b\right\} $\ and consider a
logic function $f$ over $X$. This is equivalent to the classical register
state $|f)\equiv \psi _{f}\left( a\right) \otimes \psi _{f}\left( b\right) $%
, an element of a rank-two classical register. A basis for this register is
given by the set $\mathsf{B}\left( X\right) \equiv \;\left\{
|00),|01),|10),|11)\right\} ,$ where
\begin{eqnarray}
|00) &\equiv &|0)_{a}\otimes |0)_{b}=\left[
\begin{array}{c}
0 \\
1%
\end{array}%
\right] _{a}\otimes \left[
\begin{array}{c}
0 \\
1%
\end{array}%
\right] _{b},\;\;\;  \notag \\
|01) &\equiv &|0)_{a}\otimes |1)_{b}=\left[
\begin{array}{c}
0 \\
1%
\end{array}%
\right] _{a}\otimes \left[
\begin{array}{c}
1 \\
0%
\end{array}%
\right] _{b},
\end{eqnarray}%
and so on. Relative to this basis, the register state $|f)$ has the
expansion
\begin{eqnarray}
|f) &=&\tilde{f}\left( a\right) \tilde{f}\left( b\right) |00)+\tilde{f}%
\left( a\right) f\left( b\right) |01)+f\left( a\right) \tilde{f}\left(
b\right) |10)+f\left( a\right) f\left( b\right) |11)  \notag \\
&=&\sum_{i,j=0}^{1}f_{ij}|ij),  \label{abc}
\end{eqnarray}%
where $f_{00}\equiv \tilde{f}\left( a\right) \tilde{f}\left( b\right) $, $%
f_{01}\equiv \tilde{f}\left( a\right) f\left( b\right) $, and so on.

So far, no issues of entanglement can arise because classical registers do
not support general superposition. This is reflected by the fact that the
coefficients $f_{ij}$ in the expansion (\ref{abc}) of the state $|f)$
relative to the basis $\mathsf{B}\left( X\right) $ are not only either ones
or zeros, but satisfy the condition
\begin{equation}
f_{00}f_{11}=f_{10}f_{01},  \label{micro}
\end{equation}%
which is a necessary and sufficient condition for a bipartite quantum state
to be separable, i.e., not entangled \cite{JAROSZKIEWICZ-03A}.

A \emph{local operator} is any operator on register states which acts on
each individual site with no reference to any of the other sites. Examples
are the identity operator
\begin{equation}
\mathbb{I}_{N}\equiv \sigma _{0}^{0}\otimes \sigma _{1}^{0}\otimes \ldots
\otimes \sigma _{N-1}^{0}
\end{equation}%
and the negation operator,
\begin{equation}
\mathbb{N}_{N}\equiv \sigma _{0}^{1}\otimes \sigma _{1}^{1}\otimes \ldots
\otimes \sigma _{N-1}^{1}.
\end{equation}%
Products of local operators are also local operators.

Both classical and quantum computation requires the use of non-local
operators, the most important being the CNOT, or controlled-not, gate, which
involves two sites. It is related to the XOR or measurement gate
given in equation $\left( \ref{XOR}\right) $, the difference being that the
CNOT gate has two output bits. For the rank two register system $X$ given
above, consider the non-local operator
\begin{equation}
C_{ab}\equiv P_{a}^{0}\otimes \sigma _{b}^{0}+P_{a}^{1}\otimes \sigma
_{b}^{1}.
\end{equation}%
Its effect is to flip the state at site $b$ if the state at site $a$
answers $yes$, otherwise the state at $b$ is unaltered. The first site $a$
is known as the (information) donor and the second site $b$
is the acceptor (of information). Its action on state $|f)$ returns the
state: \begin{eqnarray}
C_{ab}|f) &=&\tilde{f}\left( a\right) \tilde{f}\left( b\right) |00)+\tilde{f}%
\left( a\right) f\left( b\right) |01)  \notag \\
&&+f\left( a\right) f\left( b\right) |10)+f\left( a\right) \tilde{f}\left(
b\right) |11).  \label{156}
\end{eqnarray}%
Now a bona fide rank-two classical register state is necessarily separable,
i.e., satisfies condition $\left( \ref{micro}\right) $. By considering all
four possible values of the coefficients $f\left( a\right) $ $f\left(
b\right) $ etc., it can be readily proved that (\ref{156}) satisfies this
condition, so that it is a legitimate classical register state. However, it
is not possible to rearrange (\ref{156}) as it stands to look like a
separable state without running the risk of a formal division by zero.
Explicit separability at this point therefore requires a knowledge of the
specific logical outcomes (i.e., truth values) $f\left( a\right) $ and $%
f\left( b\right) $. Hence the result of the CNOT operation has a degree of
contextuality; although we can be sure in advance that the result of a CNOT
operation is in general a classical register state over a rank-two system,
and therefore is separable, we cannot explicitly write it out as such until
we know the details of the initial state it acts on.

There are two other properties of the CNOT operation which will be important
in any extension of our work to quantum field theory. First, the CNOT
operation implies a temporal ordering, in that it asks a question of the
state of the donor site $a$ and only then gives a response in terms of doing
something to the acceptor site $b$. This is underlined by the above
mentioned contextuality. Any question $Q$ and answer $A$ involves a natural
ordering, $\left( Q,A\right) $, which defines what can be called a logical
or computational arrow of time. With the $CNOT$ gate, it is not possible to
actually change the answer at site $b$ before we have asked the question of
site $a$ and received an answer.

The other important property of the CNOT gate $C_{ab}$ is its involvement of
two distinct bits in a non-trivial way which distinguishes it from local
operations. This non-locality is the analogue of the spatial derivative
operator in conventional quantum field theory, which is the continuum
version of a nearest neighbour interaction and without which there would be
no spacetime dynamics as such. In computational terms, the CNOT gate allows
the transmission of information over the register.

The CNOT gate is reversible, $C_{ab}$ being its own inverse. The transpose
CNOT gate $C_{ab}^{T}$ is defined by
\begin{equation}
C_{ab}^{T}\equiv C_{ba}\equiv \sigma _{a}^{0}\otimes P_{b}^{0}+\sigma
_{a}^{1}\otimes P_{b}^{1},
\end{equation}%
which merely reverses the roles of acceptor and donor, but still operates
forwards in time.

In our discussion of the bosonic oscillator, below, it turns out that the
CNOT gate does not occur on its own in general but is used to build up more
complex computational gates. A particularly important case is the transpose
gate $T_{ab}$, defined by
\begin{equation}
T_{ab}\equiv C_{ab}C_{ab}^{T}C_{ab}=C_{ab}^{T}C_{ab}C_{ab}^{T}.
\end{equation}%
The transpose gate has the effect of switching the logic values between two
sites:
\begin{equation}
T_{ab}\psi _{f}\left( a\right) \otimes \psi _{f}\left( b\right) =\left[
\begin{array}{c}
f\left( b\right) \\
\tilde{f}\left( b\right)%
\end{array}%
\right] _{a}\otimes \left[
\begin{array}{c}
f\left( a\right) \\
\tilde{f}\left( a\right)%
\end{array}%
\right] _{b}.
\end{equation}%
It may be represented in the form
\begin{eqnarray}
T_{ab} &=&\frac{1}{2}\sum_{\mu =0}^{3}\sigma _{a}^{\mu }\otimes \sigma
_{b}^{\mu }  \notag \\
&=&P_{a}^{0}\otimes P_{b}^{0}+P_{a}^{1}\otimes P_{b}^{1}+A_{a}\otimes
A_{b}^{+}+A_{a}^{+}\otimes A_{b}.
\end{eqnarray}%
Hence we may write
\begin{equation}
A_{a}\otimes A_{b}^{+}+A_{a}^{+}\otimes A_{b}=T_{ab}-P_{a}^{0}\otimes
P_{b}^{0}-P_{a}^{1}\otimes P_{b}^{1},  \label{adg}
\end{equation}%
which we shall use in our study of the quantized bosonic oscillator.

\section{Quantization}

Up to this point all quantities have involved only real numbers, which is a
requirement in classical logic and classical physics in general. When we
quantize, however, we normally introduce complex numbers into the formalism,
such as in the expression $\left[ \hat{p},\hat{x}\right] =-i\hslash $ for
the commutator of the particle observables of momentum and position.
Quantization of the classical register means regarding the classical bit
states $|1)$ and $|0)$ as the basis vectors of a complex Hilbert space known
as a \emph{quantum bit}, or \emph{qubit}. Arbitrary linear complex
combinations of these basis vectors are now permitted. The price paid for
this is that we immediately lose the interpretation of the components of
qubits in terms of classical truth-values. On the other hand, we gain a
contextual probabilistic interpretation. Unlike a classical state, a quantum
state has physical meaning only within the context of the experiment chosen
to test it.

Once we have quantized the theory, we can consider variants of result (\ref%
{adg}), arising from the fact that now we can make arbitrary unitary
transformations of the basis vectors. In particular, at a given site,
consider a non-classical change in the standard qubit basis of the form
\begin{equation}
|0)\rightarrow |0^{\prime })\equiv e^{-i\alpha }|0),\;\;\;|1)\rightarrow
|1^{\prime })\equiv e^{-i\beta }|1),
\end{equation}%
where $\alpha $ and $\beta $ are real phases. This change may be represented
by the action of the unitary operator
\begin{equation}
U\left( \alpha ,\beta \right) \equiv \left[
\begin{array}{cc}
e^{-i\alpha } & 0 \\
0 & e^{-i\beta }%
\end{array}%
\right]
\end{equation}%
acting on the qubit. Under this transformation, any operator $O$ acting on
the qubit transforms according to the rule
\begin{equation}
O\rightarrow O^{\prime }\equiv UOU^{+}.
\end{equation}%
We find%
\begin{equation}
\begin{array}{ccccl}
A & \rightarrow & A^{\prime } & = & e^{i\left( \alpha -\beta \right) }A \\
A^{+} & \rightarrow & A^{\prime +} & = & e^{i\left( \beta -\alpha \right)
}A^{+} \\
\sigma ^{1} & \rightarrow & \sigma ^{\prime 1} & = & \cos \left( \alpha
-\beta \right) \sigma ^{1}+\sin \left( \alpha -\beta \right) \sigma ^{2} \\
\sigma ^{2} & \rightarrow & \sigma ^{\prime 2} & = & \sin \left(\beta -
\alpha \right) \sigma ^{1}+\cos \left( \alpha -\beta \right) \sigma ^{2},%
\end{array}%
\end{equation}
the other elementary qubit operators remaining unchanged. This
transformation preserves the rules given in Table 1.

For a two qubit system, consider the combined transformation
\begin{equation}
U_{ab}\equiv U_{a}\left( \alpha ,\beta \right) \otimes U_{b}\left( \gamma
,\delta \right) ,
\end{equation}%
where the phases are changed independently. Then the $CNOT$ gate $C_{ab}$
transforms according to the rule
\begin{eqnarray}
C_{ab}\rightarrow C_{ab}^{\prime } &\equiv &U_{ab}C_{ab}U_{ab}^{+}  \notag \\
&=&P_{a}^{0}\otimes \sigma _{b}^{0}+P_{a}^{1}\otimes \left\{ \cos \left(
\gamma -\delta \right) \sigma _{b}^{1}+\sin \left( \gamma -\delta \right)
\sigma ^{2}\right\} .
\end{eqnarray}%
The transpose gate $T_{ab}$ transforms according to the rule
\begin{eqnarray}
T_{ab}\rightarrow T_{ab}\left( \theta \right) &\equiv &U_{ab}T_{ab}U_{ab}^{+}
\notag \\
&=&P_{a}^{0}P_{b}^{0}+P_{a}^{1}P_{b}^{1}+e^{i\theta
}A_{a}A_{b}^{+}+e^{-i\theta }A_{a}^{+}A_{b}.  \label{trans}
\end{eqnarray}%
where $\theta \equiv \alpha -\beta +\gamma -\delta $. This result will be
used in our discussion of coherent states. We find

\begin{equation}
T_{ab}\left( \theta \right) \psi _{f}\left( a\right) \otimes \psi _{f}\left(
b\right) =\left[
\begin{array}{c}
f\left( b\right) \\
e^{i\theta }\tilde{f}\left( b\right)%
\end{array}%
\right] _{a}\otimes \left[
\begin{array}{c}
f\left( a\right) \\
e^{-i\theta }\tilde{f}\left( a\right)%
\end{array}%
\right] _{b},
\end{equation}%
which switches the information between the two sites in a non-classical way
whenever $\theta $ not a multiple of 2$\pi .$ For example, choosing $\theta
=\pi /2$ we find the useful result
\begin{equation}
i\left( A_{a}A_{b}^{+}-A_{a}^{+}A_{b}\right) =T_{ab}\left( \pi /2\right)
-P_{a}^{0}P_{b}^{0}-P_{a}^{1}P_{b}^{1},  \label{145}
\end{equation}%
which will be used in our discussion of the momentum operator for the
quantized bosonic oscillator.

\section{Quantum computation}

In 1982, Feynman discussed the possibility of simulating physics in terms of
quantum computation \cite{FEYNMAN-82}. In 1995 \cite{BARENCO-95}, it was
shown that any quantum computation could be built up from local unitary
operations and CNOT gates alone. This suggests that it should be possible to
realize Feynman's vision and reinterpret all quantum theories in terms of
local unitary operations and CNOT gates.

We now show how we can rewrite the quantized bosonic oscillator in quantum
computational terms, i.e., via local operations and CNOT gates. The first
step is to discuss the bosonic quantum register, which forms the matrix in
which we set the quantized oscillator.

\section{The bosonic quantum register}

Let $\mathcal{R}^{\infty }$ be an infinite-rank classical bit register,
i.e., $\mathcal{R}^{\infty }$ is a collection of an infinite number of
classical bits, each labelled by a distinct non-negative integer $n\in
\lbrack 0,\infty )$. We shall call this register a \emph{bosonic }quantum
register to indicate that it has infinite rank. Each bit has two possible
states and therefore $\mathcal{R}^{\infty }$ contains an infinite number of
classical states, each of which is of the form
\begin{equation}
|i_{0}i_{1}i_{2}\ldots )=|i_{0})_{0}\otimes |i_{1})_{1}\otimes
|i_{2})_{2}\otimes \ldots ,\;\;\;\;\;i_{n}=0\text{ or }1\text{ for each }%
n=0,1,2,\ldots
\end{equation}%
For example,
\begin{equation}
|01101\ldots )\equiv |0)_{0}\otimes |1)_{1}\otimes |1)_{2}\otimes
|0)_{3}\otimes |1)_{4}\otimes \ldots
\end{equation}%
Clearly, each classical register state $|i_{0}i_{1}i_{3}\ldots )$
corresponds to a unique binary sequence $S:i_{0},i_{1},i_{2},\ldots $
consisting of an infinite string of ones and zeros. The connection with
logic functions is as follows. Given a classical state in $\mathcal{R}%
^{\infty }$, we ask a classical logic question at each bit. If the answer
for the $n^{th}$ bit is \textquotedblleft \emph{yes}" then the corresponding
sequence element $s_{n}\in S$ is one, otherwise it is a zero. For every
state in this register, we have to ask an infinite number of questions to
establish its corresponding sequence. A given sequence corresponds to a
unique logic function over the register.

As a set, $\mathcal{R}^{\infty }$ is non-denumerable, i.e., it is not
possible to assign a unique integer to each of its states. This creates a
potential problem when we come to quantization, because the Hilbert space $%
\mathcal{H}^{\infty}$ corresponding to $\mathcal{R}^{\infty }$ is an
infinite tensor product and such Hilbert spaces are always non-separable
\cite{STREATER+WIGHTMAN:64} (i.e., have no countable basis). This is at odds
with the fact that the quantized bosonic oscillator has a complete set of
eigenstates of the Hamiltonian which is countable. Fortunately, for the
harmonic oscillator, we can restrict our attention to a set of very special
operators (called \emph{bosonic operators}) over the quantum register, so
that in physical applications, non-separability can be avoided \cite%
{STREATER+WIGHTMAN:64,KLAUDER+SUDARSHAN:68}.

To understand how this comes about, we first classify each state in $%
\mathcal{R}^{\infty }$ as one of three possible types. Two of these types
form countable sets whilst the third type is non-denumerable. These types
correspond, roughly speaking, to the integers, rationals, and irrationals
respectively, which can be seen by the following heuristic arguments.

The first type, the set of all \emph{finite countable} states in $\mathcal{R}%
^{\infty }$, consists of states associated with binary sequences which
consist of zeros after some given finite element $J$, which depends on the
sequence. For example, the state $|11010000000\ldots )$ is finite countable (%
$J=4$) whereas the state corresponding to the infinitely recurring sequence $%
10101010\ldots $ is not. For a finite countable state $|i_{0}i_{1}i_{2}%
\ldots i_{J-1}000\ldots )$ the computational map $\left( \ref{smap}\right) $
maps this state to the integer%
\begin{equation}
i_{0}+2i_{1}+2^{2}i_{2}+\ldots +2^{J-1}i_{J-1}.
\end{equation}

The second type, the \emph{recurring sequence} states in $\mathcal{R}%
^{\infty }$, consists of those sequences which would be finite countable
sequences but for the fact that the infinite string of zeros after $J$ is
replaced by some non-trivial recurring binary sequence. Recurring sequence
states cannot be classified by finite integers using the computational map $%
\left( \ref{smap}\right) $. However, we can use another map, defined by%
\begin{equation}
|i_{0}i_{1}i_{2}\ldots )\rightarrow i_{0}+\dfrac{i_{1}}{2^{1}}+\dfrac{i_{2}}{%
2^{2}}+\ldots .
\end{equation}%
to map such states into the interval $\left[ 0,2\right] $. We shall call
this the \emph{continuum map}. It is easy to see that in fact, \emph{all}
states in $\mathcal{R}^{\infty }$ can be mapped into the interval $\left[ 0,2%
\right] $ via the continuum map. The void state $|000\ldots )$ maps into
zero whilst the fully occupied state $|111\ldots )$ maps into $2$. All other
states necessarily map into the open interval $\left( 0,2\right) $.

It is not hard to see that finite countable states and recurring sequence
states map into the rationals via the continuum map, but not into a one-to
one way. For example, the finite countable state $|1000\ldots )$ maps into
the number $1$ by the continuum map, which is also the value mapped from the
recurring sequence state $|01111111\ldots )$.

The problem with non-denumerability arises because of the existence of the
third type of infinite binary sequence. This consists of all those binary
sequences which are not recurring, such as $1110111101111100000000011\ldots $%
, an example based on the successive digits in the decimal representation of
$\pi $. There are infinitely many such sequences and they cannot be counted,
as they correspond to the irrationals, which is easy to prove.

Our conclusion is, therefore, that states in $\mathcal{R}^{\infty }$ cannot
be classified by the integers. Instead, we may use the sequence
corresponding to each state as an index, viz, if $S$ is the binary sequence $%
S\equiv \left\{ s_{0},s_{1},s_{2},\ldots \right\} $ then the corresponding
state $|S)$ is given uniquely by the expression%
\begin{equation}
|S)\equiv |s_{0})_{0}\otimes |s_{1})_{1}\otimes |s_{2})_{2}\otimes \ldots =%
\overset{\infty }{\underset{i=0}{\otimes }}|s_{i})_{i}.
\end{equation}%
Given two such register state $|S)$, $|T)$, their inner product $(S|T)$ is
defined in the obvious way, viz%
\begin{equation}
(S|T)=\left\{ \overset{\infty }{\underset{i=0}{\otimes }}{}_{i}(s_{i}|\right%
\} \left\{ \overset{\infty }{\underset{j=0}{\otimes }}|t_{j})_{j} \right\} =%
\overset{\infty }{\underset{i=0}{\Pi }}(s_{i}|t_{i})_{i}.
\end{equation}%
This takes value unity if and only if the sequences $S$ and $T$ are
identical, otherwise it is zero.

Quantization amounts to taking the set of all such states as a basis $%
\mathsf{B}$ for a non-separable Hilbert space $\mathcal{H}^{\infty }$.
States in $\mathcal{H}^{\infty }$ will be called \emph{quantum register
states} and are of the form%
\begin{equation}
|\Psi )=\sum_{S}\Psi \left( S\right) |S),
\end{equation}%
where the summation is over all possible infinite binary sequences and the
coefficients $\Psi \left( S\right) $ are complex. The Hilbert space inner
product is defined in the obvious way, viz%
\begin{equation}
(\Psi |\Phi )=\sum_{S}\Psi ^{\ast }\left( S\right) \Phi \left( S\right) .
\end{equation}%
\qquad \qquad

As discussed above, finite countable sequences can be mapped into the
integers via the computational map $\left( \ref{smap}\right) $. For a
sequence $S$ which maps into integer $n$, we can use the notation $|n)$
rather than $|S)$ to denote the corresponding quantum register state.

\section{The quantized bosonic oscillator}

The quantum register states we need to represent the quantized bosonic
oscillator form a subset of the finite countable states. To identify this
subset, we need to filter out of the set of all finite countable states
those states which are redundant.

To do this, we first define the \emph{bosonic projection operators}:
\begin{equation}
\mathbb{P}_{n}\equiv \left\{ \overset{\infty }{\underset{i\neq n}{\otimes }}%
P_{i}^{0}\right\} \otimes P_{n}^{1}=P_{0}^{0}\otimes P_{1}^{0}\otimes \ldots
P_{n-1}^{0}\otimes P_{n}^{1}\otimes P_{n-1}^{0}\otimes \ldots ,\ \ \
n=0,1,2,\ldots ,
\end{equation}%
where the superscripts label the different qubit operators discussed in
Table $1$ and the subscripts denote sites in the register. Each bosonic
projection operator $\mathbb{P}_{n}$ defines a one-dimensional Hilbert space
with basis $|2^{n})$. Eigenstates of these operators with eigenvalue $+1$
will be called \emph{bosonic eigenstates}.

Next, we define the \emph{bosonic identity }operator%
\begin{equation}
\mathbb{I}_{B}\equiv \sum_{n=0}^{\infty }\mathbb{P}_{n}.
\end{equation}%
This operator satisfies the idempotency condition required of any projection
operator, viz,%
\begin{equation}
\mathbb{I}_{B}\mathbb{I}_{B}=\mathbb{I}_{B}
\end{equation}%
and plays the role of a \textquotedblleft bosonic filter", passing only
those states and operators associated with the quantum register which have
the desired properties associated with the harmonic oscillator.

\

\noindent \textbf{Definition: }\ A quantum register operator $\mathbb{O}$ is
\emph{bosonic} if and only if it commutes with the bosonic identity, i.e.,
\begin{equation}
\left[ \mathbb{O},\mathbb{I}_{B}\right] =0.
\end{equation}

\noindent \textbf{Definition: }A \emph{bosonic state} is defined to be any
vector in the infinite rank quantum register $\mathcal{H}^{\infty }$ which
is an eigenstate of $\mathbb{I}_{B}$ with eigenvalue $+1$. All other states
in $\mathcal{H}^{\infty }$ will be referred to as \emph{transbosonic}.

\

Examples of transbosonic states are the void state $|0)$ and all those
finite countable elements $|n)$ of the basis $\mathsf{B}$ where $n$ is not
some power of two. In fact, almost all elements in the quantum register are
transbosonic. It can be readily verified that linear superpositions of
bosonic states are always bosonic states, whilst linear superpositions of
any transbosonic state with any other state in the register is always
transbosonic.

The importance of the bosonic operators is that when they are applied to
bosonic states, the result is always a bosonic state, which can be easily
proved.

\

We now in a position to discuss how we map the standard quantum oscillator
onto the quantum register.

For the quantum oscillator, the most important operators are the ladder
operators $a$ and $a^{+}$. We can use them to discuss coherent (Glauber)
states and the more traditional one-particle states. We shall focus on the
latter now and deal with the former towards the end of this paper.

Given the generic form for the classical one-dimensional oscillator
Hamiltonian
\begin{equation}
H=\frac{_{1}}{^{2}}\alpha ^{2}p^{2}+\frac{_{1}}{^{2}}\beta ^{2}x^{2},
\label{111}
\end{equation}%
where $\alpha ,\beta $ are positive constants, we define the ladder
operators in the usual way:
\begin{eqnarray}
\hat{a} &\equiv &\beta \hat{x}+i\alpha \hat{p},  \notag \\
\hat{a}^{+} &\equiv &\beta \hat{x}-i\alpha \hat{p},
\end{eqnarray}%
where $\hat{x},\hat{p}$ are the usual bosonic position and momentum
operators satisfying the standard Weyl-Heisenberg algebra
\begin{equation}
\left[ \hat{p},\hat{x}\right] =-i\hbar .
\end{equation}%
Then we find
\begin{equation}
\left[ \hat{a},\hat{a}^{+}\right] =2\varepsilon \hat{I},
\end{equation}%
where $\varepsilon \equiv \alpha \beta \hbar $ $\equiv \omega \hbar $ has
the physical dimensions of an energy and $\hat{I}$ is the identity operator
in the standard oscillator Hilbert space $\mathcal{H}$.

Using standard arguments, it is easy to show that the ladder operators have
the representations
\begin{eqnarray}
\hat{a} &=&\sum_{n=0}^{\infty }|n\rangle \sqrt{\left( n+1\right)
2\varepsilon }\langle n+1|,  \notag \\
\hat{a}^{+} &=&\sum_{n=0}^{\infty }|n+1\rangle \sqrt{\left( n+1\right)
2\varepsilon }\langle n|,
\end{eqnarray}%
where the states $|n\rangle ,\;n=1,2,3,\ldots $ are the usual orthonormal
excited states of the ground state $|0\rangle $. These states are identified
one-to-one with the bosonic states $|2^{n})$ discussed above, viz
\begin{equation}
|n\rangle \leftrightarrow |2^{n}),\ \ \ n=0,1,2,\ldots .
\end{equation}

To find a quantum register representation of the ladder operators, we first
introduce some auxiliary notation. We define%
\begin{equation}
\mathbb{P}^{0}\equiv \overset{\infty }{\underset{i=0}{\otimes }}P_{i}^{0},\
\ \ \mathbb{A}_{n}\equiv \left\{ \overset{\infty }{\underset{i\neq n}{%
\otimes }}\sigma _{i}^{0}\right\} \otimes A_{n},\ \ \ \mathbb{A}%
_{n}^{+}\equiv \left\{ \overset{\infty }{\underset{i\neq n}{\otimes }}\sigma
_{i}^{0}\right\} \otimes A_{n}^{+}.
\end{equation}%
None of these operators is bosonic as can be readily proved. In fact, the
operators $\mathbb{A}_{n}^{+}$ can be used to generate bosonic states from
the void state (which is transbosonic). Specifically, we have%
\begin{equation}
|2^{n})=\mathbb{A}_{n}^{+}|0)\text{.}
\end{equation}%
With these definitions we construct the operators
\begin{equation}
\mathbb{B}_{n}^{+}\equiv \mathbb{A}_{n}^{+}\mathbb{P}^{0}\mathbb{A}_{n+1},\
\ \ \mathbb{B}_{n}\equiv \mathbb{A}_{n+1}^{+}\mathbb{P}^{0}\mathbb{A}_{n}.
\end{equation}%
Remarkably, these operators are bosonic, as can be readily proved from the
fact that%
\begin{equation}
\mathbb{A}_{n}\mathbb{I}_{B}=\mathbb{A}_{n}\text{.}
\end{equation}%
To simplify the notation, we shall henceforth leave out the implied tensor
product symbols $\otimes $. Then we find%
\begin{eqnarray}
\mathbb{B}_{n} &\equiv &\left\{ \overset{\infty }{\underset{i\neq n,n+1}{%
\otimes }}P_{i}^{0}\right\} A_{n}A_{n+1}^{+}=P_{0}^{0}P_{1}^{0}\ldots
P_{n-1}^{0}A_{n}A_{n+1}^{+}P_{n+2}^{0}\ldots  \notag \\
\mathbb{B}_{n}^{+} &\equiv &\left\{ \overset{\infty }{\underset{i\neq n,n+1}{%
\otimes }}P_{i}^{0}\right\} A_{n}^{+}A_{n+1}=P_{0}^{0}P_{1}^{0}\ldots
P_{n-1}^{0}A_{n}^{+}A_{n+1}P_{n+2}^{0}\ldots \;\;\;n=0,1,2,\ldots ,  \notag
\\
&&
\end{eqnarray}%
where the symbols and superscripts on the right hand side refer to Table $1$
and the subscripts label different qubits. These operators satisfy the
relations
\begin{eqnarray}
\mathbb{B}_{n}^{+}\mathbb{B}_{m} &=&\delta _{nm}\mathbb{P}_{n+1}  \notag \\
\mathbb{B}_{n}\mathbb{B}_{m}^{+} &=&\delta _{nm}\mathbb{P}_{n}.
\end{eqnarray}%
Then the ladder operators take the quantum register representation
\begin{eqnarray}
a &\leftrightarrow &a_{B}\equiv \sum_{n=0}^{\infty }\sqrt{\left( n+1\right)
2\varepsilon }\mathbb{B}_{n},  \notag \\
a^{+} &\leftrightarrow &a_{B}^{+}\equiv \sum_{n=0}^{\infty }\sqrt{\left(
n+1\right) 2\varepsilon }\mathbb{B}_{n}^{+}.
\end{eqnarray}%
These operators are bosonic and have the commutation relation%
\begin{equation}
\left[ a_{B},a_{B}^{+}\right] =2\varepsilon \mathbb{I}_{B}.
\end{equation}%
Because these register ladder operators commute with the bosonic identity $%
\mathbb{I}_{B},$ any states that they create from bosonic states are also
bosonic. We find for example
\begin{equation}
|n\rangle \equiv \frac{\left( a^{+}\right) ^{n}}{\sqrt{n!\left( 2\varepsilon
\right) ^{n}}}|0\rangle \leftrightarrow \frac{\left( a_{B}^{+}\right) ^{n}}{%
\sqrt{n!\left( 2\varepsilon \right) ^{n}}}|1)=|2^{n}),\ \ \ n=0,1,2,\ldots
\end{equation}%
Note that $a_{B}$ annihilates the bosonic ground state $|1)$, giving the
zero vector $0$ in the quantum register $\mathcal{H}^{\infty }$, not the
void state $|0)$.

The quantized bosonic Hamiltonian (\ref{111}) has the quantum register
representation
\begin{equation}
\hat{H}\leftrightarrow \mathbb{H}_{B}\equiv \frac{_{1}}{^{2}}a_{B}^{+}a_{B}+%
\frac{_{1}}{^{2}}\varepsilon \mathbb{I}_{B}=\sum_{n=0}^{\infty }\left( n+%
\frac{_{1}}{^{2}}\right) \varepsilon \mathbb{P}_{n}.
\end{equation}%
It commutes with the bosonic identity $\mathbb{I}_{B}$ and therefore, any
states in the quantum register which start off bosonic at initial time
remain bosonic, if they evolve under this Hamiltonian.

\section{Quantum computation}

It is not immediately apparent from the quantum register representation of
the Hamiltonian in which way computational operations are involved in
bosonic oscillator dynamics, particularly given that the number states $%
|n\rangle $ $\leftrightarrow |2^{n})$ are eigenstates of the Hamiltonian and
therefore have trivial time dependence. To see how bosonic dynamics involves
quantum register computation, recall that the position and momentum
operators in the standard theory are given by
\begin{equation}
\hat{x}=\frac{1}{2\beta }\left[ a^{+}+a\right] ,\;\;\;\hat{p}=\frac{i}{%
2\alpha }\left[ a^{+}-a\right] .
\end{equation}%
We represent them in the quantum register description by the operators%
\begin{equation}
\hat{x}\leftrightarrow \hat{x}_{B}\equiv \dfrac{1}{2\beta }\left[
a_{B}^{+}+a_{B}\right] ,\ \ \ \hat{p}\leftrightarrow \hat{p}_{B}\equiv
\dfrac{i}{2\alpha }\left[ a_{B}^{+}-a_{B}\right] .
\end{equation}%
These operators are bosonic, by virtue of the fact that the operators $%
a_{B},a_{B}^{+}$ are bosonic. Then we find%
\begin{equation}
\left[ \hat{x}_{B},\hat{p}_{B}\right] =i\hslash \mathbb{I}_{B}.
\end{equation}%
In the quantum register representation, these operators take the forms
\begin{eqnarray}
\hat{x}_{B} &=&\frac{1}{2\beta }\sum_{n=0}^{\infty }\sqrt{\left( n+1\right)
2\varepsilon }\left\{ \mathbb{B}_{n}^{+}+\mathbb{B}_{n}\right\}  \notag \\
&=&\frac{1}{2\beta }\sum_{n=0}^{\infty }\sqrt{\left( n+1\right) 2\varepsilon
}\left\{ \overset{\infty }{\underset{j\neq n,n+1}{\otimes }}%
P_{j}^{0}\right\} \left\{ A_{n}A_{n+1}^{+}+A_{n}^{+}A_{n+1}\right\}  \notag
\\
&=&\frac{1}{2\beta }\sum_{n=0}^{\infty }\sqrt{\left( n+1\right) 2\varepsilon
}\left\{ \overset{\infty }{\underset{j\neq n,n+1}{\otimes }}%
P_{j}^{0}\right\} \left\{ T_{n,n+1}\left( 0\right)
-P_{n}^{0}P_{n+1}^{0}-P_{n}^{1}P_{n+1}^{1}\right\} ,  \notag \\
&&
\end{eqnarray}%
and

\begin{eqnarray}
\hat{p}_{B} &=&\frac{i}{2\alpha }\sum_{n=0}^{\infty }\sqrt{\left( n+1\right)
2\varepsilon }\left\{ \mathbb{B}_{n}^{+}-\mathbb{B}_{n}\right\}  \notag \\
&=&\frac{1}{2\alpha }\sum_{n=0}^{\infty }\sqrt{\left( n+1\right)
2\varepsilon }\left\{ \overset{\infty }{\underset{j\neq n,n+1}{\otimes }}%
P_{j}^{0}\right\} \left\{ iA_{n}A_{n+1}^{+}-iA_{n}^{+}A_{n+1}\right\}  \notag
\\
&=&\frac{1}{2\alpha }\sum_{n=0}^{\infty }\sqrt{\left( n+1\right)
2\varepsilon }\left\{ \overset{\infty }{\underset{j\neq n,n+1}{\otimes }}%
P_{j}^{0}\right\} \left\{ T_{n,n+1}\left( \pi /2\right)
-P_{n}^{0}P_{n+1}^{0}-P_{n}^{1}P_{n+1}^{1}\right\} ,  \notag \\
&&
\end{eqnarray}%
where $T_{n,n+1}\left( 0\right) $ and $T_{n,n+1}\left( \pi /2\right) $ are
specific cases of the operator $\left( \ref{trans}\right) $.

We note now that the operators $\hat{x}_{B}$ and $\hat{p}_{B}$ are bosonic
operators, i.e., they commute with the bosonic identity, $\mathbb{I}_{B}$.
Moreover, these operators act only on bosonic states, i.e., eigenstates of $%
\mathbb{I}_{B}$ with eigenvalue $\emph{+1}$. Suppose we have such a bosonic
state $|\Psi )$. Then
\begin{eqnarray}
\hat{x}_{B}|\Psi ) &=&\hat{x}_{B}\mathbb{I}_{B}|\Psi )  \notag \\
&=&\frac{1}{2\beta }\sum_{n=0}^{\infty }\sqrt{\left( n+1\right) 2\varepsilon
}\;\left\{ \overset{\infty }{\underset{j\neq n,n+1}{\otimes }}%
P_{i}^{0}\right\} T_{n,n+1}\left( 0\right) |\Psi ),
\end{eqnarray}%
because the operators
\begin{equation}
\mathbb{P}^{0}\equiv \otimes _{j=0}^{\infty }P_{j}^{0},\;\;\;\;\;\left\{
\overset{\infty }{\underset{j\neq n,n+1}{\otimes }}P_{j}^{0}\right\}
P_{n}^{1}P_{n+1}^{1},\;\;\;n=0,1,2,\ldots
\end{equation}%
annihilate the bosonic identity. This can be seen from the properties of
Table $1$, which hold for every qubit in the quantum register. The same
argument applies to the momentum operator and we find
\begin{equation}
\hat{p}_{B}|\Psi )=\frac{1}{2\alpha }\sum_{n=0}^{\infty }\sqrt{\left(
n+1\right) 2\varepsilon }\left\{ \overset{\infty }{\underset{j\neq n,n+1}{%
\otimes }}P_{j}^{0}\right\} T_{n,n+1}\left( \pi /2\right) |\Psi ),
\end{equation}%
if $|\Psi )$ is a bosonic state. \ Because all physically relevant operators
representing observables, such as the Hamiltonian $\mathbb{H}_{B}$, can be
constructed out the $\hat{x}_{B}$ and $\hat{p}_{B}$, this establishes the
basic result, which is that the bosonic oscillator can indeed be discussed
in terms of quantum computation. The $P_{i}^{0}$ operators are local unitary
operators and the transpose operators $T_{n,n+1}\left( \theta \right) $ are
essentially CNOT gates. This agrees with the results of Barenco \emph{et al}
\cite{BARENCO-95}, where it was shown that any quantum computation could be
built up from local unitary operations and CNOT gates alone.

\section{Coherent states}

In quantum optics, photon states occur frequently in certain types,
depending on the preparation procedures. Three important types are single
photon states, coherent states and thermalized states. Each of these have
their individual experimentally observable properties which allow the
experimentalist to determine their nature from observations of ensembles of
such states. Coherent states are important in quantum optics because they
are essentially the sort of states created by lasers. They are also
important in particle physics.

A coherent state $|z\rangle $, $z\in \mathbb{C}$ is a linear superposition
of single particle states with the property that%
\begin{equation}
\hat{a}|z\rangle =z\sqrt{2\varepsilon }|z\rangle ,\ \ \ z\in \mathbb{C}.
\end{equation}%
With our conventions, we find%
\begin{equation}
|z\rangle =e^{-\tfrac{1}{2}|z|^{2}}\sum_{n=0}^{\infty }\dfrac{z^{n}}{\sqrt{n!%
}}|n\rangle .
\end{equation}%
Such states can also be represented by the action of the shift operator%
\begin{equation}
D\left( z\right) \equiv e^{-\tfrac{1}{2}|z|^{2}}e^{z\hat{a}^{+}/\sqrt{%
2\varepsilon }}e^{-z^{\ast }\hat{a}/\sqrt{2\varepsilon }}=\exp \left\{
\dfrac{z\hat{a}^{+}-z^{\ast }\hat{a}}{\sqrt{2\varepsilon }}\right\}
\end{equation}%
acting on the ground state $|0\rangle $ \cite{KLAUDER+SUDARSHAN:68}.

The quantum register version $\mathbb{D}\left( z\right) $ of this operator
is given by%
\begin{equation}
\mathbb{D}\left( z\right) =\exp \left\{ \sum_{n=0}^{\infty }\sqrt{\left(
n+1\right) }\left( z\mathbb{B}_{n}^{+}-z^{\ast }\mathbb{B}_{n}\right)
\right\}
\end{equation}%
acting on the bosonic state $|2^{0})$. If we write%
\begin{equation*}
z\equiv ire^{i\theta }
\end{equation*}%
then we find%
\begin{eqnarray}
\mathbb{D}\left( z\right) &=&\exp \left\{ ir\sum_{n=0}^{\infty }\sqrt{\left(
n+1\right) }\left( e^{i\theta }\mathbb{B}_{n}^{+}-e^{-i\theta }\mathbb{B}%
_{n}\right) \right\}  \notag \\
&=&\exp \left\{ ir\sum_{n=0}^{\infty }\left[ \overset{\infty }{\underset{%
j\neq n,n+1}{\otimes }}P_{j}^{0}\right] \sqrt{\left( n+1\right) }\left\{
T_{n,n+1}\left( \theta \right)
-P_{n}^{0}P_{n+1}^{0}-P_{n}^{1}P_{n+1}^{1}\right\} \right\}  \notag \\
&&
\end{eqnarray}%
using $\left( \ref{trans}\right) .$ Since this operator acts on $|2^{0}),$
which is an eigenstate of the bosonic identity $\mathbb{I}_{B}$, we may
replace it by the expression%
\begin{equation}
\mathbb{D}\left( z\right) \rightarrow \mathbb{D}^{\prime }\left( z\right)
\equiv \exp \left\{ ir\sum_{n=0}^{\infty }\left[ \overset{\infty }{\underset{%
j\neq n,n+1}{\otimes }}P_{j}^{0}\right] \sqrt{\left( n+1\right) }\left\{
T_{n,n+1}\left( \theta \right) \right\} \right\} ,
\end{equation}%
which displays the quantum computational structure of coherent states.

\subsection{Concluding remarks}

The view explored by Feynman, Fredkin and others during the early
nineteen eighties that the universe might be describable in
computational terms is an attractive idea which gave rise to the current
intense interest in quantum computers. What we have shown in this paper is
that this idea certainly works for a system as useful as the bosonic
oscillator, which finds many applications throughout physics. We expect that
a similar, though inevitably more detailed, analysis should hold for all
free particle quantum field theories, as these can be regarded as systems of
coupled oscillators.

Our result immediately raises interesting questions about more complicated
coupled systems, such as the anharmonic oscillator, and by implication,
interacting quantum field theories. The quantum register approach rests on
the properties of very particular operators, as given in Table $1$, and
these cannot be altered. Therefore, any new interactions can come about only
because of novel \textquotedblleft non-local" interactions between register
sites. Essentially, a quantum register imposes its own rules on the
formalism. Work is in hand to understand how to use this approach for more
complex systems.

\section{Acknowledgements}

J.Ridgway-Taylor thanks the EPSRC for a grant.

\end{document}